\documentclass[prl,twocolumn,showpacs]{revtex4}
\usepackage{amsmath}
\usepackage{amssymb}
\usepackage{float}
\usepackage{bm,times}
\usepackage{graphicx}
\newcommand{\be}{\begin{equation}}
\newcommand{\ee}{\end{equation}}
\newcommand{\bea}{\begin{eqnarray}}
\newcommand{\eea}{\end{eqnarray}}
\newcommand{\ek}{\epsilon_{\mathbf{k}}}

\newcommand{\Ek}{E_{\mathbf{k}}}

\newcommand{\phik}{\varphi_{\mathbf{k}}}
\newcommand{\phikq}{\varphi_{{\mathbf{k}}-{\mathbf{q}}/2}}

\newcommand{\sumk}{\sum_{\mathbf{k}}}

\newcommand{\sumq}{\sum_{\mathbf{q}}}

\newcommand{\Omegaq}{\Omega_{\mathbf{q}}}

\begin{document}

\title{Population of closed-channel molecules in trapped Fermi gases
  with broad Feshbach resonances}

\date{\today}

\author{Qijin Chen}
\author{K. Levin}

\affiliation{James Franck Institute and Department of Physics,
University of Chicago, Chicago, Illinois 60637}

\begin{abstract}

  We compute the fraction of closed-channel molecules in trapped
  atomic Fermi gases, over the entire range of accessible fields and
  temperatures. We use a two-channel model of BCS--Bose-Einstein
  condensation (BEC) crossover theory at general temperature $T$, and
  show that this fraction provides a measure of the $T$ dependent
  pairing gap.  Our calculations, containing no free parameters, are
  in good quantitative agreement with recent low $T$ measurements in
  $^6$Li. We present readily testable predictions for the dependencies
  of the closed-channel fraction on temperature and Fermi momentum.

\end{abstract}

\pacs{03.75.Hh, 03.75.Ss, 74.20.-z \hfill \textsf{Phys. Rev. Lett. \textbf{95},
  260406 (2005)}}

\maketitle


The superfluid phases of ultracold, trapped atomic Fermi gases present
an exciting opportunity to study how superfluidity is changed as the
system evolves from the weak attraction limit of
Bardeen-Cooper-Schrieffer (BCS) state to the strong attraction
Bose-Einstein condensation (BEC) regime \cite{Leggett}.  Feshbach
resonances arise when a bound state (``closed channel molecules") of a
two-body spin-singlet potential lies near the continuum onset of a
spin-triplet scattering state (``open channel"). As a consequence of an
applied magnetic field $B$, proximity to these resonances can profoundly
change the two-body ($s$-wave) scattering lengths $a$.  The well studied
Feshbach resonances in both $^6$Li and $^{40}$K are very wide as a
function of $B$, compared to the Fermi energy.  Moreover, the focus is
on unitary scattering where $a$ diverges.  For these reasons one often
resorts to a one-channel version of BCS-BEC crossover theory
\cite{Leggett,Strinati}, although attention has also been paid to the
two-channel variant \cite{Holland,JS2a,Griffin,Stoof2,Pethick} for wide
as well as narrow Feshbach resonances.

In recent experiments \cite{Hulet4} on the broad resonance in $^6$Li at
low temperatures, $T$, a laser drives transitions between dressed
molecules (hybridized with open channel atom pairs) and an excited
molecular singlet, in the process removing excited atoms from the trap.
The exponentially decreasing number of remaining trapped atoms as a
function of probe duration time is used to determine that fraction
$\cal{Z}$ of dressed molecules/pairs corresponding to the closed
channel.  Here we present the analogous theoretical calculations of
$\cal{Z}$, as a function of $a$ and for general $T$. We demonstrate very
good agreement with experiment, and provide predictions for future
finite $T$ experiments.  The small size of this fraction means that for
many properties the two-channel model reduces to the single-channel
description.  We emphasize however, that the two-channel model contains
more information making it capable of addressing a wider range of
physical systems and experiments, and of testing the fundamentals of
BCS-BEC crossover theory.

Our theoretical approach reduces to the generalized Leggett-BCS theory
\cite{Leggett} at $T =0$.  At finite $T$, the noncondensed as well as
condensed contributions enter in the appropriate combinations, so that a
future measurement of the total closed-channel fraction will provide a
measure of the $T$-dependent pairing gap $\Delta(T)$. This gap is, in
general, distinct from the superfluid order parameter,
$\tilde{\Delta}_{sc}(T)$, except at $T=0$.  Finally, we present results
for the closed-channel fraction as a function of the global Fermi
wavevector $k_F$, at different fields $B$.  At unitarity, the dependence
is a simple proportionality.  When two-channel physics is important, as
it must be in the determination of the closed-channel fraction, then the
energy scale which characterizes the width of the Feshbach resonance
also enters into the problem.  In this way this fraction is
non-universal and assumes different values for $^6$Li and $^{40}$K.

We begin with a discussion of the superfluid phase.  A gas of Fermi
atoms in the presence of a Feshbach resonance contains open-channel
fermionic atoms as well as closed-channel molecules.  Because of
coupling $g$ between the open and closed channels the three propagators
for the open-channel fermion pairs [$t_{pg}(Q)$], the closed-channel
molecules [$D(Q)$] and the single (open-channel) fermion states [$G(K)$]
are all highly interconnected.  Here and throughout, we use a
four-momentum notation: $K\equiv (i\omega_n, \mathbf{k})$ and $Q\equiv
(i\Omega_m, \mathbf{q})$, $\sum_K \equiv k_B T \sum_n\sumk$, and their
analytical continuation, $i\omega_n \rightarrow \omega+i0^+, i\Omega_m
\rightarrow \Omega + i 0^+$, where $\omega_n = (2 n+1) \pi k_B T/\hbar$,
$\Omega_n = 2 n \pi k_BT/\hbar$ are the odd and even Matsubara
frequencies.  The $T$-matrix scheme we employ to treat these coupled
propagators is derived from the equations of motion for the Green's
functions, and, importantly, it naturally leads to the self-consistency
conditions of the standard $T=0$ mean field theory \cite{Leggett}.  It
can be shown \cite{JS3} that the pairs are described by the pair
susceptibility $\chi(Q) = \sum_K G_0(Q-K)G(K)\phikq^2$ where $G$ depends
on a BCS-like self energy $ \Sigma(K) \approx -\Delta^2 G_0(-K)\phik^2
$.  Here $G_0(K)$ is the noninteracting fermion Green's function.
Throughout this paper $\phik \equiv \exp\{-k^2/2k_0^2\}$ introduces a
momentum cutoff, where $k_0$, represents the inverse range of
interaction, which is assumed infinite for a contact interaction.

Our Hamiltonian is the standard boson-fermion model
\cite{Griffin,Holland,ourreview} in which there are only fermion-boson
and fermion-fermion interactions.  Central to our analysis at finite $T$
are noncondensed pairs \cite{JS3} which we characterize by a $T$-matrix
$t_{pg}(Q) = U_{eff}(Q)/[1 + U_{eff}(Q) \chi(Q)]$, where $U_{eff}$ is
the effective pairing interaction which involves the direct two-body
interaction $U$ as well as virtual excitation processes associated with
the Feshbach resonance \cite{Griffin,JS3}.  At small $Q$, $t_{pg}$ can
be expanded as
\begin{equation}
t_{pg}(Q) \approx \frac { Z^{-1}}{\Omega - \Omega_q +\mu_{pair} + i \Gamma_Q}.
\label{eq:expandt}
\end{equation}
The parameters appearing in Eq.~\ref{eq:expandt} are discussed in more
detail in Ref. ~\cite{ourreview}.  Here $Z^{-1}$ is a residue and $
\Omega_q$ the pair dispersion.  The latter parameter as well as the
effective pair chemical potential $\mu_{pair}$ both depend on the
important, but unknown, gap parameter $\Delta$ through the fermion self
energy $\Sigma$. The decay width $\Gamma_Q$ is negligibly small for
small $Q$ below $T_c$.  We caution here that the inverse residue ``Z''
appearing in Eq.~(\ref{eq:expandt}) is not the same as the quantity
$\cal{Z}$ used in Ref.~\cite{Hulet4}.  More technical details about the
various residues can be found in Ref.~\cite{ourreview}.

To determine $\Delta$, one imposes the BEC-like constraint $\mu_{pair}
=0 $ which yields $t_{pg}^{-1}(Q\rightarrow 0) = 0=U_{eff}^{-1}(0) +
\chi(0)$, i.e.,
\begin{equation}
U_{eff}^{-1}(0) + \sumk \frac{1-2f(\Ek)}{2\Ek} \phik^2 = 0 \,,
\label{eq:Gap}
\end{equation}
so that $\Delta$ formally satisfies the usual BCS gap equation with
quasiparticle dispersion $\Ek = \sqrt{(\ek - \mu)^2 + \Delta^2\phik^2}$,
where $\ek=\hbar^2k^2/2m$ is the fermion kinetic energy, $\mu$ is the
fermionic chemical potential, and $f(x)$ is the Fermi distribution
function.
  
Physically, one should view $\Delta$ as reflecting the presence of
bosonic degrees of freedom. In the fermionic regime ($\mu>0$), it
represents the energy required to break the pairs.  It can be seen
\cite{JS3} that $\Delta$ contains contributions from both noncondensed
and condensed pairs, whose densities are proportional to
$\Delta^2_{pg}(T)$ and $\tilde{\Delta}_{sc}^2(T)$, respectively.  One
has a constraint on the \textit{number of pairs} \cite{ourreview}
which can be viewed as analogous to the usual BEC number constraint
\begin{equation}
\Delta^2(T) = \tilde{\Delta}_{sc}^2(T) + \Delta_{pg}^2(T) \,.
\label{eq:3}
\end{equation}
The total contribution of noncondensed pairs is readily computed in
terms of $\Delta$ via
\begin{equation}
 \Delta_{pg}^2 \equiv -  \sum_Q t_{pg}(Q) \,.
 \label{eq:PG}
 \end{equation}
In analogy with the standard derivation of BEC, one can then compute
the number of condensed pairs associated with $\tilde{\Delta}_{sc}$.

Fermions are the fundamental particles in this system, so that their
chemical potential $\mu$ is determined from the number conservation
constraint
\begin{equation}
n = n_f + 2 n_{b0} + 2n_b  \equiv n_f + 2 n_b^{tot}  \,,
\label{eq:Number}
\end{equation}
where $n_{b0}$ and $n_b$ represent the density of condensed and
noncondensed closed-channel molecules, respectively, $n_b^{tot}$ is their
sum, and $n_f = 2\sum_K G(K)$ is the atomic density associated with the
open-channel fermions.  Here
\begin{equation}
n_b = -\sum_Q D(Q)  \approx Z_b \sumq b(\Omegaq -\mu_{pair}) \,,
\label{eq:nb}
\end{equation}
where $b(x)$ is the Bose distribution function.  The renormalized
closed-channel molecule propagator $D(Q)$ is given by the same equation
as Eq.~(\ref{eq:expandt}) with a different residue $Z^{-1} \rightarrow
Z_b$.  We compute $\Delta$ (and $\mu$) via Eqs.~(\ref{eq:Gap}) and
(\ref{eq:Number}), to determine the contribution from the condensate
$\tilde{\Delta}_{sc}$ via Eqs.~(\ref{eq:3}) and (\ref{eq:PG}). Note that
at $T=0$, $\Delta_{pg}=0$ so that all pairs are condensed as is
consistent with the mean-field BCS-Leggett ground state.

At $T=0$, a non-zero excitation gap for all $\bf{k}$ constrains $n_f$ so
that one can identify our closed-channel fraction $2n_b^{tot}/n =
2n_{b0}/n$ with the quantity $\cal{Z}$ in experiment.  In this picture
the ground state consists of $n/2$ pairs with fraction $2n_{b0}/n$ in
the closed channel.  
A short range attractive interaction (appropriate for cold Fermi gases),
leads to a picture in which all fermions are paired, since the
attraction extends to the entire Fermi sphere. Importantly, this is 
consistent, at least at unitarity, with the observed exponential
behavior of the remaining total particle number in the trap found in
Ref.~\onlinecite{Hulet4}.  Thus, we presume here that the ground state
has no unpaired fermions.

The interaction parameters $U$ and $g$ which appear in the Hamiltonian
can be related to their experimental counterparts $U_0$ and $g_0$. The
latter are, in turn, determined by the open-channel background
scattering length $a_{bg}$ and the Feshbach resonance width $W$: $U_0
= 4\pi a_{bg}\hbar^2 / m = 8\pi k_F a_{bg} (E_F/k_F^3)$ and $g_0^2 =
|U_0W|$.  Here $E_F= \hbar^2k_F^2 /2m$ is the global noninteracting
Fermi energy.
From the Lippmann-Schwinger equation \cite{Kokkelmans} we obtain 
$1/U_0 = 1/U - 1/U_c \,,$
where $1/U_c \equiv -\sumk (\phik^2/2\ek)$ is the value of the
interaction corresponding to unitary scattering ($a = \infty$).
Similarly, we define $U^*\equiv 4\pi a^*\hbar^2 /m$ and the ``experimental''
magnetic detuning $\nu_0$ such that 
\begin{equation}
\label{eq:U*}
U^* = U_0 + \frac{g^2_0}{2\mu-\nu_0} \,,\quad
\frac{1}{U^*} = \frac{1}{U_{eff}} - \frac{1}{U_c} \,.
\end{equation}
As in Ref.~\onlinecite{Kokkelmans}, defining $\Gamma=1/(1+U_0/U_c)$, one
has $U=\Gamma U_0$, $g = \Gamma g_0$ and $\nu = \nu_0 - \Gamma g_0^2
/U_c$.
In order to make contact with experiment, it is convenient
to define a two-body counterpart
of $U^*$ in Eq.~(\ref{eq:U*}):
\begin{equation}
U_{2B}\equiv  4\pi a\hbar^2 /m = U_0 - g^2_0 / \nu_0 \,,
\label{eq:U2B}
\end{equation}
which diverges at $\nu_0 =0$. Here
$\nu_0 = (B-B_0) \Delta\mu^0$, where $B_0$ is the resonance field, and
$\Delta\mu^0 =2\mu_B$ for $^6$Li is the difference in the magnetic
moment between open-channel pairs and the closed-channel molecules
\cite{Chin95}. Here $\mu_B$ is the Bohr magneton.
For the wide Feshbach resonances in $^{40}$K and $^6$Li, the
dimensionless parameters $1/k_Fa$ and $1/k_Fa^*$ are very close to each
other for the magnetic fields addressed in Ref.~\cite{Hulet4}.

It is useful, now to rewrite the closed-channel density in terms of
more physically accessible parameters.  Building on
Refs.~\cite{Griffin,Holland}, one can show \cite{JS3} $n_{b0}\propto
\tilde{\Delta}_{sc}^2 $.  It follows from our 
$T \neq 0$ formalism \cite{ourreview} that when a condensate is 
present  
\begin{equation}
  n_{b0} 
  = Z_g \tilde{\Delta}_{sc}^2 \,,\quad n_{b} 
  = Z_g \Delta_{pg}^2 =
\frac{\Delta_{pg}^2} {g_0^2}\left(1-\frac{a_{bg}}{a^*}\right)^2 \,,
\label{eq:nb0}
\end{equation}
where we have used some simple algebra to rewrite $Z_g\equiv
g^2/[(2\mu-\nu)U+g^2]^2$. Importantly, in this way we can conclude from
Eq.~(\ref{eq:3}) that the total closed-channel contribution $n_b^{tot}
= Z_g \Delta^2$.  The simplicity of this last result reflects the fact
that $D(Q)$ and $t_{pg}(Q)$ [of Eq.~(\ref{eq:expandt})] share the same
denominator. The relative probabilities for the (hybridized) pairs to
live in the closed and open channels are essentially fixed for all $T
\leq T_c$ and given by $Z_b = Z_g/Z$ and $ 1 - Z_b$, respectively.

To represent the trap, we use the local density approximation (LDA) by
replacing $\mu\rightarrow \mu(r) \equiv \mu -V(r)$.  Here $\mu$ is the
global chemical potential and $V(r) = m\omega^2 r^2/2 $ for a harmonic
trap with angular frequency $\omega$.  We solve Eqs.~(\ref{eq:PG}) and
(\ref{eq:Gap}) at each $r$ for given $\mu$ and then self-consistently
adjust $\mu$ to satisfy the total number constraint $N=\int
\mathrm{d}^3r \,n(r)$.  We define $N_{b0} \equiv \int \mathrm{d}^3r\,
n_{b0}(r)$ and similarly for the noncondensed molecules $N_{b}$, and the
total $N_b^{tot}=N_{b0}+N_b$.  Here $N_{b0}$ represents the trap average
of the superfluid order parameter.  Moreover, this trap average is
proportional to the value of the order parameter at the center so that $
N_{b0}/N \propto \tilde{\Delta}^2_{sc}(r=0)$. Thus, this full
two-channel calculation provides a theoretical underpinning for the
simple interpretation, provided in Ref.~\cite{Hulet4}, of their
experiments.

\begin{figure}
\centerline{\includegraphics[clip,width=3.2in]{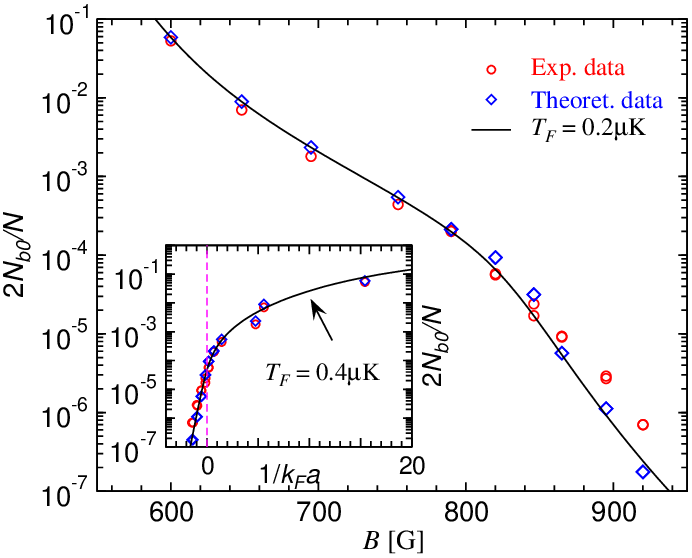}}
\caption{(Color online) Calculated closed-channel fraction
  $2N_{b}^{tot}/N =2N_{b0}/N$ at $T=0$ as a function of magnetic field
  $B$ (main figure) at $T_F=0.2\mu$K for $^6$Li in a harmonic trap. The
  inset shows $2N_{b0}/N$ as a function of $1/k_Fa$ (black curve) at
  $T_F=0.4\mu$K.  The red circles are experimental measurements from
  Ref.~\cite{Hulet4}, and the blue diamonds are theoretically calculated
  data using specific values of $T_F$ from experiment
  \cite{HuletPrivate}, which vary between 0.18-0.662~$\mu$K.  }
\label{fig:kFa}
\end{figure}

In the unitary regime where $|a_{bg}/a^*| \ll 1$, Eq.~(\ref{eq:nb0})
implies $n_{b0} \approx \tilde{\Delta}^2_{sc} /g_0^2$, at each point $r$.
Therefore, we can deduce that at unitarity
\begin{equation}
  \frac{2N_{b0}}{N\;} =\int \mathrm{d}^3r
  \frac{\tilde{\Delta}_{sc}^2(r)}{ g_0^2N} \propto  \frac{\int \mathrm{d}^3r
    \tilde{\Delta}^2_{sc}(r)/E_F^2}{\int \mathrm{d}^3r\, n(r)/n(0)}\,
  \frac{\sqrt{E_F}}{ g_0^2} \propto 
  \sqrt{T_F}\,.
\label{eq:12}
\end{equation}
At unitarity, $2N_{b0}/N$ scales with $k_F$. Here we have used the
relationship $N=n(0)\int \mathrm{d}^3r\, [n(r)/n(0)] $ and $n(0) \propto
E_F^{3/2}$. The fraction in front of $\sqrt{E_F}/ g_0^2$ is
dimensionless and independent of $T_F$.

In Fig.~\ref{fig:kFa}, we plot the calculated molecular fraction
$2N_{b0}/N$ at $T=0$ for $^6$Li (black curve and blue diamonds) as a
function of the magnetic field $B$ as compared with experimental data
(red circles) from Ref.~\cite{Hulet4}. The experimental values for $T_F$
vary from one data point to another.  Our calculations at each field
value use as input the specific experimental value for $T_F$, plotted as
blue diamonds, in the figure and inset.  However for the continuous
curve plotted in the main figure, we use $T_F= 0.2\mu$K which better
represents the values in the BCS regime \cite{HuletPrivate}.  In the
inset we used $T_F = 0.4 \mu$K, since this best reflects the average
value over the entire range of data points.  These slightly different
choices for $T_F$ reflect the fact that the two plots have different
horizontal axes, and, therefore, amplify small errors in different ways.
As will be shown below, in the BCS regime the results are particularly
sensitive to $T_F$.  We take the parameters for the Feshbach resonance
from Ref.~\cite{GrimmJuliennea}: $W=300$~G and a field-dependent $a_{bg}
= a_{bg}^0/[1+\alpha(B-B_0)]$, where $a^0_{bg}=-1405a_0$, $\alpha =
0.0004$~G$^{-1}$, $B_0=834.15$~G and $a_0=0.529$\AA\ is the Bohr radius.
For $T_F=0.4\mu$K, this uniquely determines $U_0=-5.90 E_F/k_F^3$ and
$g_0=-771 E_F/k_F^{3/2}$ at unitarity.

\begin{figure}
\centerline{\includegraphics[clip,width=3.in]{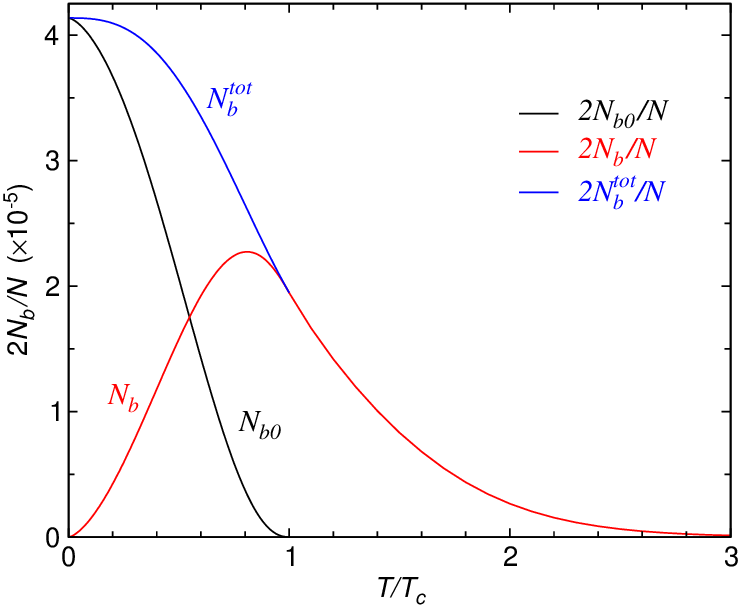}}
\caption{(Color online) Closed-channel fraction as a function of $T/T_c$
  at unitarity for $T_F = 0.4\mu$K for $^6$Li in a harmonic trap. The
  black, red, and blue curves are the condensed ($2N_{b0}/N$),
  noncondensed ($2N_b/N$) and total ($2N_b^{tot}/N$) fractions,
  respectively. Here $T_c=0.273T_F$. In the BEC regime $2N_b^{tot}/N$ is
  relatively $T$ independent below $T_c$.}
\label{fig:T} 
\end{figure}

In Fig.~\ref{fig:kFa} the
roughly a factor of 3 
difference between theory and experiment in the BCS regime may in part
be associated with the observed deviation from an exponential time
dependence \cite{Hulet4}.  Due to the extremely small gap ($\lesssim T$)
at these high fields, unpaired fermions may be thermally excited.
However, it should be noted that this particular thermal ``correction''
alone would be in the wrong direction (see below). This may point to the
need for a more systematic treatment of dynamical effects in this
regime.
Overall, agreement with the data is quantitatively very good
\cite{Hulet4} everywhere but in the BCS regime;
there are no adjustable parameters \cite{footnoteonk0-Li}.

We turn to the $T$ dependence of the closed-channel fraction, which is
plotted at unitarity in Fig.~\ref{fig:T}.  This fraction was shown above
to be proportional to the pairing gap squared $\Delta^2(T)$.  The
condensed fraction $2N_{b0}/N$ (black curve) decreases as $T$ increases
from zero and vanishes at $T_c$. At the same time, the noncondensed
fraction $2N_b/N$ (red) increases from zero, as determined by
Eq.~(\ref{eq:nb}). It has a maximum slightly below $T_c$.  The total
fraction $2N_b^{tot}/N$ (blue) decreases monotonically with $T$. Note
that it decreases very slowly at low $T$. This justifies our comparison
in Fig.~\ref{fig:kFa} between the $T=0$ calculations and the low $T$
measurements. Using Eqs.~(\ref{eq:nb0}) and (\ref{eq:12}), and in
conjunction with the temperature insensitivity of the various residues
($Z_g$ and $Z_b$), one can measure the $T$ dependence of the pairing
gap, presuming that $T$ is determined from isentropic sweeps
\cite{ChenThermo}.  The sensitivity of the total fraction to temperature
increases with field.

\begin{figure}
\centerline{\includegraphics[clip,width=3in]{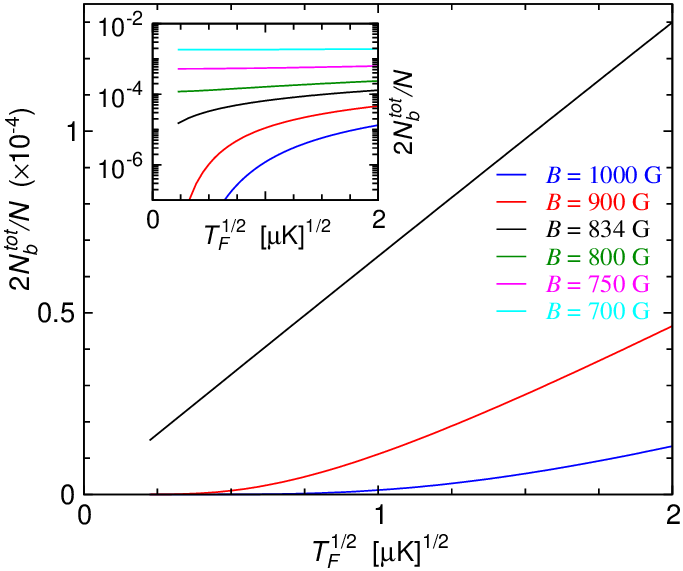}}
\caption{(Color online) Predicted closed-channel fraction $2N_{b0}/N$ at
  $T=0$ as a function of $\sqrt{T_F} \propto k_F$ at different fields
  for $^6$Li in a harmonic trap. At unitarity ($B=834$G, black line),
  $2N_{b0}/N \propto k_F$. On the BCS side ($B=900$G, red, and
  $B=1000$G, blue), this becomes a higher power law. On the BEC side
  ($B<834$G, see the inset, $B$ increases from top to bottom), the
  fraction becomes less sensitive to $T_F$, and is determined by
  two-body physics in the deep BEC limit.  
}
\label{fig:kF}
\end{figure}

Next we explore the relationship between the closed-channel fraction and
the Fermi temperature $T_F$.  In Fig.~\ref{fig:kF}, we plot $2N_{b0}/N$
at $T=0$ as a function of $\sqrt{T_F} \propto k_F$ for different
magnetic fields, from the weak pairing BCS ($B=1000$G, blue curve) to
the strong pairing BEC ($B=700$G, cyan curve). It can be seen that the
plots reflect our earlier theoretical prediction that $2N_{b0}/N \propto
k_F$ at unitarity. As the field increases, $2N_{b0}/N$ varies as a
higher power of $k_F$. In contrast, as the field decreases in the BEC
regime, it becomes less sensitive to $k_F$.  For $^6$Li, it approaches
unity in the deep BEC limit, where it is dominated by two-body physics.
More generally, $2N_{b0}/N$ increases faster (slower) on the BCS (BEC)
side than the simple proportionality found at unitarity, as is confirmed
by Fig.~\ref{fig:kF}.

In summary, we find that the wider is the resonance the more readily the
closed-channel molecules decay into the open channel so that the steady
state molecular fraction remains small, as observed experimentally.  We
have shown here that measurements of the closed-channel fraction will
complement other techniques for obtaining the pairing gap.  While broad
Feshbach resonances often lead to universal behavior at unitarity
\cite{Thomas}, a single dimensionless parameter $k_Fa$ is inadequate for
determining quantities such as $2N_{b0}/N$, which is intrinsically
associated with two-channel physics.


We thank R. Hulet for sharing data, and J.E. Thomas, C.A. Regal, M.
Greiner, and C. Chin for discussions and the OCTS workshop.  This work
was supported by NSF-MRSEC Grant No.~DMR-0213745 and by the Institute
for Theoretical Sciences and DOE, No.  W-31-109-ENG-38 (QC).


\textit{Note added}. --- After submission of this Letter, two papers
appeared on the $T=0$ closed-channel fraction within a homogeneous
system based on a different model for the ground state \cite{Romans} and
on a limiting case \cite{Javanainen} of the present theory in which
direct interfermion interactions and closed-channel self-energy) are
ignored.


\bibliographystyle{apsrev}

\end{document}